# How to Detect and Measure the AI Dangers to Democracy

Giulia Sandri, Université libre de Bruxelles
Claudio Novelli, DEC, Yale University

**Abstract.** Research on artificial intelligence and democracy has grown quickly over the last decade. A shared conclusion in this literature is that AI does not create new democratic problems so much as it makes old ones worse. We now see this across information ecosystems, in elections, and in public administration. However, despite growing evidence, we lack a clear way to prioritize risks in this area, compare them across domains, and identify where democratic control is most likely to break down. So, our problem is: How can we systematize the problems that AI systems pose to democratic processes? This paper argues that principal–agent theory may fit the task. In many phases of democratic systems, principals delegate key functions to AI systems and their providers without really being able to monitor how these systems operate or the outputs they produce. Treating AI as a delegation problem helps identify accountability gaps and other governance failures. Most importantly, as we shall illustrate, it provides metrics for empirical assessments of AI impact on democracy. As a second analytical element, we draw on the NIST AI Risk Management Framework (NIST 2023) and its seven characteristics of trustworthy AI, which supply substantive criteria for evaluating delegated tasks: the principal–agent perspective identifies where democratic delegation is most exposed, while the trustworthiness criteria specify what principals should assess when agents act on their behalf. Operationalized across the three domains through measurable indicators and domain-specific trustworthiness criteria, we propose an analytical framework that centers on institutional assessability as the central condition for democratic control over AI. However, we stress that how severe a harm is, and how much risk is acceptable, are evaluative judgments that current methodologies neither acknowledge nor operationalize. This becomes acute when such evaluative judgments are (silently) delegated to private vendors. We identify this as a strong limitation left for future work.

## 1. Introduction

Over the past decade, academic research on the relationship between artificial intelligence and democracy has expanded rapidly, reflecting growing concern about how automated decision-making, large-scale data analytics, and generative models affect and reshape political processes (Suleyman, 2023; Narayanan & Kapoor, 2024; Novelli & Sandri, 2025). Scholars increasingly emphasize that AI systems do not merely augment existing institutions' ordinary functions with new tools, but recalibrate fundamental democratic functions: e.g., representation, participation, and accountability (Kreps & Kriner, 2023; Fink-Hafner, 2025). Empirical studies document democratic gains for governments, such as improved administrative capacity and enhanced citizen engagement, and substantial risks, ranging from algorithmic discrimination and information manipulation to the concentration of technological power in a few state or corporate actors, which are often foreign vendors (Innerarity 2024; Coeckelbergh 2025; König 2025; Persily & Tucker 2026). A unifying theme across the emerging literature is that AI's political effects are mediated by institutional design, regulatory choices, and broader geopolitical dynamics, making its democratic implications deeply dependent on context (Jungherr 2023).

For this reason, it is important to develop a framework for assessing the overall impact of AI adoption in democratic processes. Although empirical evidence is expanding across different areas of contemporary politics (König 2025; Persily & Tucker 2026), we still lack a systematic way to prioritize risks, compare them across domains, and identify where democratic control is most likely to break down.

To address this, we propose an analytical framework for assessing (a) what it means for AI systems to exhibit agency in political settings and (b) how hybrid human–AI arrangements reallocate responsibility and decision-making authority across key democratic domains (information integrity, elections and participation, and public service delivery).

The framework recognizes that when public administrations or democratic institutions adopt AI, they are delegating tasks, and, in practice, portions of authority, to interconnected socio-technical systems. Such delegation can create new capabilities, but it also introduces new forms of opacity, dependence, and risk.

We maintain that the progressive adoption of AI in politics and democratic processes give rise to a new kind of principal-agent problem. Principal-agent problems refer to the difficulty that arises when "*agents act strategically to promote their own interests at the expense of the principal's goals*" (Cofone & Strandburg 2026: 6), because the latter cannot perfectly monitor the agent's actions. For this reason, our framework is grounded on the principal-agent theory (PAT), which analyses the conflict of interest that arises when one party (the principal) delegates decision-making to another (the agent) who possesses private information and whose interests may not align with the principal's (Grossman & Hart 1983; Cofone & Strandburg 2026). This approach, as shown from previous work by Jungherr et al. (2026) and the EU Joint Research Centre (Almeida et al. 2026), provides the most pertinent model for developing an analytical framework to assess the domains in which democratic control is more likely to falter when AI systems are adopted and when they consequently redistribute responsibility and decision-making authority.

As a second analytical element, we draw on the NIST AI Risk Management Framework (NIST 2023), which articulates seven characteristics of trustworthy AI: valid and reliable, safe, secure and resilient, accountable and transparent, explainable and interpretable, privacy-enhanced, and fair with harmful bias managed. These characteristics provide substantive criteria against which AI systems delegated by democratic principals can be evaluated, and they complement the principal–agent perspective by specifying what acceptable performance by an AI agent should look like in the political domain.

## 2. The democratic context and the nature of AI

AI's expansion coincides with a sustained wave of global democratic backsliding (Nord et al., 2025). While AI is not a primary driver of this trajectory, it intersects with existing political dynamics in ways that can amplify democratic vulnerabilities, undermine institutional resilience, and destabilise the functioning of political systems. Of the main risks posed by AI to politics and society identified in recent studies (Saeri et al. 2026; Slattery et al. 2026), most of the priority risks are human institutional failure modes that predate computing by centuries, such as discrimination, inequality, fraud, power centralization, and governance failure. Al did not generate them, but it accelerates them and alters the scale. The genuinely

AI-specific risks mostly concern security vulnerabilities, dangerous capabilities, robustness failures, misalignment, interpretability gaps, and possibly loss of consensus reality at scale.

This study examines the risks raised by the growing adoption of AI systems for democratic processes, encompassing both generative AI and algorithmic technologies such as recommender systems and content-curation algorithms. We adopt here a functional definition of AI, viewing it as "*machine-based systems that is capable of influencing the environment by producing an output (predictions, recommendations or decisions) for a given set of objectives. It uses machine and/or human-based data and inputs to (i) perceive real and/or virtual environments; (ii) abstract these perceptions into models through analysis in an automated manner (e.g., with machine learning), or manually; and (iii) use model inference to formulate options for outcomes. AI systems are designed to operate with varying levels of autonomy" (*Estevez Almenzar et al., 2022: 12).

The spreading adoption of AI, particularly generative AI, in political processes significantly affects information ecosystems, electoral competition, and public administration in European states (Ojanen et al. 2026; Almeida et al. 2026).

Regarding information ecosystems, AI restructures three processes: information flows, information quality, and the economics of news production (Jungherr & Schroeder, 2023). First, AI shifts environments from search-based discovery toward generative synthesis, with users receiving AI-generated summaries from Generative AI tools rather than locating sources themselves (Makhortykh et al. 2024, 2025; Gillespie 2024). This concentrates gatekeeping power among a few technology companies and risks reducing exposure to diverse viewpoints (Magin et al. 2021). Unlike search engines, LLMs produce fluent summaries that obscure sources, can generate ungrounded claims, and discourage critical engagement with plural sources (Jungherr & Schroeder 2023).

Second, Generative AI lowers the cost and skill needed to manufacture credible looking but inauthentic text, images, and video (Budak et al. 2024; Jungherr 2025), fuelling manipulation concerns (Olejnik 2025; Vaccari & Chadwick 2020). Measuring the individual impact of AI-generated disinformation remains difficult (Farooq & De Vreese 2025, 2026), and audiences often recognize and distrust synthetic media (Altay & Gilardi 2024; Toff & Simon 2024). The greater danger lies in the erosion of trust in journalism itself (Jungherr & Rauchfleisch 2025).

Third, AI threatens journalism's financial sustainability (Chapekis & Lieb 2025), diverting attention and ad revenue from news outlets, while licensing deals deepen dependency (Posnett 2025) and AI-assisted production blurs authorship (Cools & Diakopoulos 2026; Simon 2025).

In campaign operations, as recently demonstrated by Jungherr et al. (2026), parties increasingly use AI to manage volunteers, forecast donations, and predict voter behaviour (Foos 2024; Kruschinski et al. 2025; Novelli et al. 2024; Scherer et al. 2026), reducing costs but transferring influence to consultants and vendors (Martin et al. 2024) and potentially increasing political volatility (De Vries & Hobolt 2020; Jungherr et al. 2019).

For voter outreach, AI enables large-scale personalization through targeted ads and chatbots (Juneja 2024; Hackenburg et al. 2025; Votta et al. 2025), raising concerns about manipulation and unequal access (Karppinen et al. 2025; Ash et al. 2025; Fink-Hafner 2025). Evidence on persuasion effects is however mixed (Salvi et al. 2025; Hackenburg & Margetts 2024).

Concerning electoral integrity, deepfake fears are widespread, but systematic deceptive use by mainstream parties is rare (Foos 2024; Kapoor & Narayanan 2024; Jungherr, Rauchfleisch & Wuttke 2026). While AI can strengthen electoral integrity through more effective registration management, anomaly detection, and political finance oversight (Wolfs 2025; Power & Jeong 2026), it generates vendor dependencies and cybersecurity risks (Suleyman & Bhaskar 2025).

Generative AI in Public Administration and public service delivery affects evidence-informed policymaking, the efficiency of service delivery, and citizen inclusion (Van Noordt & Misuraca 2022; Rizk & Lindgren 2025; Tangi et al. 2026). While AI enhances the timeliness and effectiveness of the analysis of datasets on crises, climate, and migration (Pahlka 2023; Skotnicka-Zasadzień & Wolniak 2025), predictive models risk embedding bias in public decision-making. AI promises more efficient services (McDaniel & Pease 2021; Mollick 2024; Petrone 2025), while creating dependence on private providers and digital sovereignty risks (Scharre 2018).

Although AI can broaden inclusion (Landemore 2024; Jungherr & Rauchfleisch 2025), AI-assisted direct democracy may erode deliberation and trust through synthetic content, as illustrated recently by a *Los Angeles Times* investigation that identified more than 20,000 AI-generated public-comment emails on Southern California air-quality rules. AI thus "risks converting citizen-initiated mechanisms of direct democracy into engines of plebiscitarian instability" (Altman 2026).

## 3. Delegation problems and democratic control

Across all domains, AI heightens delegation dilemmas when elected or accountable principals (governments, media outlets, or parties) rely on autonomous or privately controlled systems (agents) that they cannot fully supervise. This follows the typical principal-agent framework.

Principal-agent problems occur when a "principal" (the party delegating a task) cannot fully observe or contractually enforce what the "agent" (the party doing the task) actually does (Wuttke, Rauchfleisch & Jungherr 2025). When the agent's actions are hidden or hard to verify, the agent may end up serving their own interests rather than the principal's (Grossman & Hart 1983). A typical example is a government agency hiring a contractor: such relationships often raise concerns about the quality of information shared, whether goals are aligned, who is accountable, and how much control the principal really has (Bovens 2007; Novelli, Taddeo & Floridi 2023). The principal-agent framework provides the most fitting political adaptation of the human and civic agency issue in AI adoption, and the most adequate model for measuring the risk of diminishing human decision-making authority when integrating AI into public and private sectors (OECD 2025; Moon & Boudreaux 2026).

AI systems, and their vendors, often play this agent role: they are given tasks that are hard to oversee or challenge. Elected officials, for instance, cannot fully track or steer how AI systems and their developers act on the public's behalf, which creates risks of mismatched goals and accountability gaps (Borch 2022; Phelps & Ranson 2023). The problem is further complicated by power imbalances between the people who design, build, and run these systems and the people who are affected by them, both those who use the systems directly and those who are simply subject to their decisions. Tackling this requires institutional safeguards: transparency (for example, through technical audits), the ability to contest decisions within the relevant legal frameworks (usually via redress procedures), and democratic oversight of AI-driven decisions in most relevant political and democratic processes (Pi & Proctor 2025).

This diffusion of power erodes transparency and accountability across multiple domains. In information ecosystems, platforms act as unseen gatekeepers of visibility, determining what content reaches audiences without clear oversight. In elections, parties increasingly rely on algorithmic tools whose operations they cannot fully audit, introducing a fundamental asymmetry between their use and understanding of their logic. Meanwhile, in public administration, officials may end up delegating authority to opaque systems, creating "second order" accountability gaps where responsibility becomes difficult to trace or assign (Koops, Hildebrandt & Jaquet-Chiffelle 2010). Therefore, the fundamental mechanism to observe for assessing the consequences of integrating AI in democratic processes is delegation: the ways in which democratic actors outsource decision-making, communication, and administrative functions to AI-enabled systems.

PAT thus offers a general account of delegation under conditions of information asymmetry and potential goal divergence (Eisenhardt 1989) that has long structured the analysis of democratic accountability as a chain of delegation from citizens to representatives, governments, and bureaucracies (Strøm 2000; Miller 2005), and it therefore frames AI systems and the actors who deploy them as a new class of agents inserted into that chain, and treating delegation issues as familiar pathologies of adverse selection, moral hazard, and accountability gaps in algorithmically mediated governance (Busuioc 2021).

As a complement to this delegation perspective, our framework also draws on the NIST AI Risk Management Framework (NIST 2023), which articulates seven characteristics of trustworthy AI. Whereas principal–agent theory identifies *where* democratic delegation is most vulnerable, the NIST trustworthiness criteria specify *what* democratic principals should evaluate when AI agents act on their behalf.

Understanding this mechanism through principal-agent theory and AI Risk management frameworks helps clarify how AI reshapes delegation dynamics and where democratic safeguards may erode. We illustrate how in the next section.

## 4. AI risk measurement and their limits

Research on how to measure AI risks has expanded along two largely parallel tracks. A first track, rooted in computer science, engineering, and governance, has produced principles, frameworks, and standards to make AI risks identifiable and manageable across the system lifecycle. Early surveys mapped the proliferation of AI ethics guidelines (Mittelstadt et al. 2016; Jobin, Ienca and Vayena 2019), highlighting recurring themes such as accountability, transparency, fairness, and privacy, as well as a persistent gap between high-level principles and practical metrics (Floridi and Cowls 2022). More recent work has translated these principles into operational frameworks (Sass et al. 2026; Novelli et al. 2024).

The NIST AI Risk Management Framework (NIST 2023) is among the most comprehensive: it identifies seven characteristics of trustworthy AI, namely being valid and reliable, safe, secure and resilient, accountable and transparent, explainable and interpretable, privacy-enhanced, and fair with harmful bias managed, intended to structure assessment in practice. Parallel efforts include the OECD AI Principles, the ISO/IEC 42001 management system standard, and the EU AI Act, which classifies AI systems into risk tiers tied to documentation, audit, and oversight obligations.

Empirical work on AI risk measurement has focused on three challenges that NIST itself flags as unresolved (NIST 2023). First, metrics are immature as there is little consensus on how to quantify trustworthiness characteristics, and laboratory measurements often diverge from real-world performance. Second, third-party dependencies (pre-trained models, opaque vendor systems, externally curated data) make risk attribution still challenging (Raji et al. 2020). Third, risks change across the lifecycle, with harms latent at design time surfacing during deployment as systems adapt and users repurpose tools beyond their intended use. Empirical responses include algorithmic audits, model and data documentation practices, explainability tooling, and impact assessments (Novelli, Taddeo, and Floridi 2024).

A second, more recent track examines AI risks specifically in political and democratic contexts, where measurement is complicated by the fact that the relevant harms are often relational and systemic rather than localized to a single system. Studies of information ecosystems have struggled to attribute individual-level effects to specific generative or recommender systems (Budak et al. 2024); studies of campaigns and elections have documented the spread of AI tools among political actors but have produced mixed evidence on persuasion effects (Jungherr, Rauchfleisch and Wuttke 2026); and studies of public administration have emphasized opacity and accountability gaps in algorithmic decision-making by the state (Busuioc 2021; Tangi, Müller and Janssen 2025). A common finding is that political risks emerge from interactions between AI tools, institutional design, and broader political dynamics (König 2025; Persily & Tucker 2026), making them difficult to capture with metrics designed for individual systems.

A further limitation, largely orthogonal to those discussed above, concerns the nature of risk itself. Risk science has long resisted the assumption that risk is a purely objective, mind-independent quantity to be read off a system. In the dominant decision-theoretic schema, risk is the product of the probability and the severity of an adverse event. Yet both the severity ascribed to outcomes and the threshold at which residual risk becomes "acceptable" (sometimes called risk tolerance or attitude) are evaluative determinations (Fischhoff et al. 1978; Slovic 1987; Hansson 2010).

This matters for AI risk measurement because the prevailing frameworks tend to treat trustworthiness and harm severity as if they were objective technical properties amenable to measurement, while offering little guidance to eliciting, representing, and aggregating the divergent risk attitudes and value weightings of those affected (i.e., the subjective component). The fundamental-rights impact assessment under the EU AI Act, which is a form of risk assessment, yields inconsistent, poorly comparable results precisely because the values and rights at stake are contested and vary across communities (Sass et al. 2026). Outside risk science proper, this valorial dimension is rarely acknowledged and almost never operationalized, with the result that a contested value judgment is presented as a neutral measurement.

The democratic stakes of this gap are considerable, and they connect directly to the delegation problem set out in Section 3. When acceptability risk judgments are folded into the design of an AI system and delegated to a private vendor, the underlying value choices become as opaque as the technical operations that carry them. The subjective, valorial component of risk thus complicates measurement and, even more importantly, raises the further accountability question of who is authorized to set the risk attitude on the public's behalf.

Across both tracks, four open problems motivate the framework developed in the next section: the lack of consensus on metrics for trustworthiness; the tendency to treat AI as a self-contained artifact rather than as an agent embedded in delegation chains; the significant variation of political risks across political delegation domains; and the lack of understanding of the valorial dimension of risk. The framework that follows combines the NIST trustworthy-AI characteristics with the principal–agent perspective developed in Section 3.

## 5. A Principal-Agent framework for empirical assessment of AI's impact on democracy

Building on the principal–agent perspective set out in Section 3 and the trustworthy-AI characteristics surveyed in Section 4, we now operationalize the framework. Through a principal-agent approach, we can distinguish three main domains of democratic processes in which power delegation occurs: the information ecosystems, elections, and public administration domains (in Table 1, they are "Domain of AI use"), and these need to be analyzed to gauge the effects of the adoption of AI systems on decision-making authority.

What makes AI delegation democratically relevant is not technological change per se, but the systematic transfer of discretion to agents who operate with limited transparency, weak accountability mechanisms, and informational advantages over their principals. For this reason, in each domain, we identify a principal, namely a citizen, a party, an elected official, or a civil servant, who delegates a function to an AI system or its corporate provider acting as an agent. We shall call this relationship the "Delegation Mechanism" (as per Table 1), and it will be the unit of analysis. The framework enables structured comparison of how these agency problems manifest across democratic dimensions, and how their severity varies across regulatory contexts and institutional settings.

First, within information ecosystems, we examine the delegation (mechanism) of authority from citizens and their representatives (principals) to information intermediaries (agents). Here, platform algorithms and generative AI systems act as agents that curate, synthesize, and distribute information. Key empirical indicators include:

1. *concentration ratios of information gatekeepers*, measured through market share of dominant platforms;
2. *diversity metrics for information exposure*, tracking to what extent algorithmic curation narrows or broadens the range of sources and perspectives that citizens encounter in main digital platforms;
3. and *transparency indices*, assessing whether platforms disclose algorithmic decision rules and allow independent audits (Helberger et al., 2018; Araujo et al. 2018).

Empirical research would measure changes in these indicators over time and across jurisdictions with different regulatory approaches, testing whether stronger transparency requirements or algorithmic accountability mechanisms correlate with maintained or improved information pluralism.

Second, in the elections domain, our framework focuses on the delegation mechanism by which political parties and candidates (principals) entrust AI vendors and consultants (agents) with tasks. Empirical assessment, thus, requires tracking:

1. the extent of *algorithmic intermediation in campaign operations*, quantified through surveys of party expenditures on AI tools and consultant services;
2. *information asymmetries between parties and vendors*, measured by assessing whether parties possess in-house expertise to audit algorithmic tools or must rely entirely on external providers;
3. and *electoral volatility patterns*, by testing whether AI adoption is associated with greater party-system fragmentation, the entry of new political movements, or shifts in competitive dynamics. One plausible mechanism for this to happen is that AI lowers barriers to entry for smaller, newer, and sometimes fringe actors by reducing organizational, financial, and informational disadvantages that traditionally favour established parties. As a consequence, this would enable low-resource campaigns to mobilize supporters and generate content at scale (Valentim and Dinas 2024; König 2025).

Comparative analysis across democracies could test whether variations in data protection regimes, transparency requirements for political advertising, or public funding for campaign infrastructure affect the severity of principal-agent problems in this domain.

Third, in the public administration domain, our framework examines delegation from elected officials and civil servants (principals) to AI systems and their corporate providers (agents). This requires measuring:

1. *procurement dependencies*, tracking the proportion of critical administrative functions delegated to external AI vendors versus developed through public infrastructure;
2. *degree of algorithmic opacity in decision-making*, assessed through the number of freedom of information requests, testing whether agencies can explain how AI systems reached specific decisions. This captures the extent to which decision authority remains transparent, allowing public decisions to be shaped by systems that can be audited or contested by officials, affected citizens, or oversight bodies. Other possible metrics are the type and number of formal appeal channels and the amount of technical documentation accessible to Public Administrations' staff;
3. and *capacity asymmetries*, comparing technical expertise within government agencies (in terms of numbers of technical/specialised FTEs) against the sophistication of deployed AI systems (Busuioc, 2021). Moreover, longitudinal studies could evaluate whether investments in public AI capacity, mandatory explainability requirements, or independent oversight bodies lead to a reduction in agency loss over time.

Each delegation mechanism affected by AI use crosses, at least, five core dimensions of democracy: participation, representation, competitiveness, transparency and accountability, and responsiveness. These dimensions correspond to the main democratic norms that are most often explored in the context of electoral politics and the quality of democracy (Kersting and Baldersheim 2004; Diamond and Morlino 2004; Kersting 2012; Rahat and Shapira 2017; Tomini and Sandri 2018). These norms reflect a broad scholarly consensus on the core procedural and substantive dimensions that define liberal democracy, including free and fair competition, inclusive participation, accountable and responsive institutions, and the rule of law (Dahl 1971; Norris 2012).[1] Each AI adoption domains relates to specific democracy dimensions and could lead to specific delegation and accountability failures, as illustrated in Figure 1.

**Fig. 1.** AI adoption domains, democracy dimensions and delegation failures

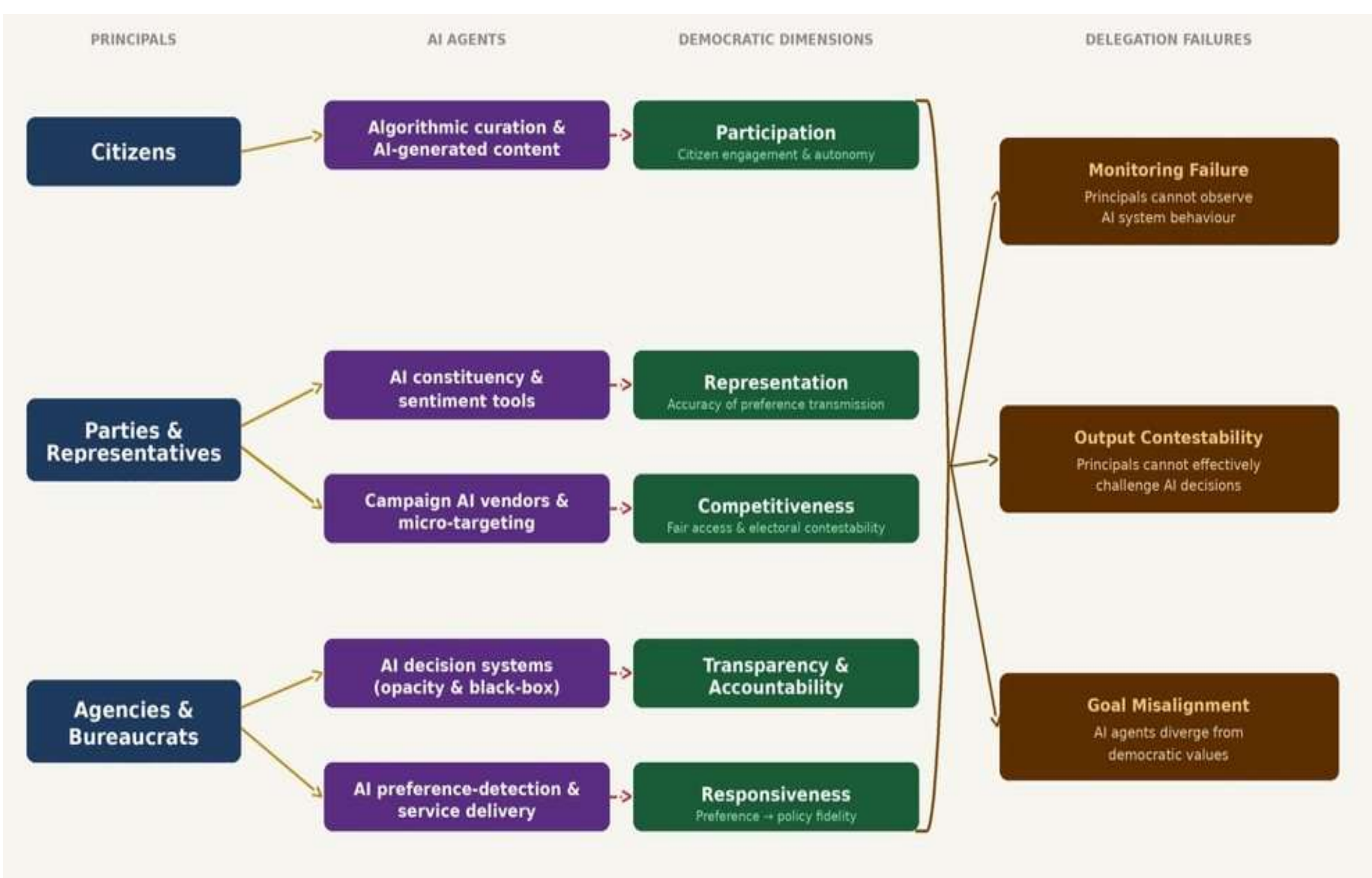

The framework specifies the theoretical mechanism through which delegation generates democratic costs within the principal-agent relationship introduced by AI; and a set of empirical indicators through which these costs can be measured and compared. The suggested indicators provide additional pertinent metrics. This last element is the framework's key methodological contribution. By grounding each dimension in observable, quantifiable indicators, ranging from algorithmic transparency indices and campaign expenditure data to freedom of information request outcomes and policy congruence measures, the framework moves beyond conceptual diagnosis to enable systematic, cross-national empirical assessment of AI's democratic impact.

---

[1] While democracy is a multidimensional concept (Lijphart 2012[1999]; Coppedge et al. 2011), not all dimensions are equally salient across all political contexts. The selection of dimensions prioritized here is therefore guided by their established centrality in comparative democratic research. Accordingly, in information ecosystems, the most affected dimension of democracy is participation; in elections, representation and competitiveness; and in public administration, transparency/accountability and responsiveness.

The framework proposes institutional assessability as the crucial mediating variable between AI deployment and democratic outcomes (Bovens et al., 2014; Heider et al. 2023; Jungherr & Schroeder 2023). Assessability entails that principals can monitor and understand decision-making processes of AI systems used in the public domain (Newman et al. 2022; Wuttke, Rauchfleisch & Jungherr 2025). The goal of the proposed framework is thus to offer an analytical tool that ultimately allows for evaluating and then improving the degree of overall assessability of AI systems deployed in the selected democratic processes. Assessability encompasses ex-ante (whether AI system design, training data, and decision rules are documented and accessible) and ex-post accountability (whether independent experts can reconstruct and verify algorithmic decisions) (Novelli, Taddeo, and Floridi 2023); and contestability (whether affected parties have effective mechanisms to challenge AI decisions through administrative or judicial review). Institutional oversight would operationalize these dimensions through regulatory compliance audits, analyzing whether jurisdictions with stronger assessability requirements demonstrate lower levels of agency loss across the three democratic domains. This is crucial for guaranteeing democratic control over AI systems deployed in public-facing domains.

**Table 1**: Impact of AI use and empirical measurement indicators across democratic dimensions

| Domain of AI Use | Delegation Mechanism | Empirical Measurement Indicators |
|---|---|---|
| **Information ecosystems** | *Citizens delegating information processing and political mobilisation to AI-driven platforms, by accepting algorithmic curation on social media and information platforms, creates structural agency problems:*<br><br>a) Platform operators and AI vendors acquire control over participation pathways without electoral mandate or formal accountability mechanisms<br><br>b) AI-generated content (chatbots, synthetic personas, automated amplification) further distorts the informational environment in which participation takes place, potentially undermining the authenticity and autonomy of citizen engagement.<br><br>c) Concentration of gatekeeping in a small number | 1. **Gatekeeper concentration**: market share of dominant platforms as a proxy for gatekeeping power (sources: digital market share data, e.g. Statcounter, Ofcom, European Digital Media Observatory (EDMO))<br>2. **Information diversity metrics**: algorithmic audit studies measuring range of sources and perspectives offered to citizens; filter-bubble indices (sources: studies from Algorithm Watch, Mozilla Foundation, EDMO)<br>3.**Platform transparency indices**: disclosure of algorithmic decision rules; availability of independent audits (sources: DSA Transparency Database, national transparency reports)<br>4.**AI-generated content prevalence**: share of political content identifiable as AI-produced or algorithmically amplified (e.g. botometer-type measures such as the CampAIgn Tracker developed by the U.Amsterdam, platform transparency reports)<br>5.**Regulatory variation**: cross-jurisdictional comparison of transparency requirements and their correlation with diversity metrics (sources: OECD.AI Policy Observatory, CEPS AI World databases) |

| Domain of AI Use | Delegation Mechanism | Empirical Measurement Indicators |
|---|---|---|
| | of dominant platforms intensifies this delegation risk. | |
| Elections | *Elected representatives relying on AI systems for constituency analysis, sentiment monitoring, and communication targeting, causes agency inversion:*<br><br>a) AI tools filter what counts as "citizen preference", potentially misrepresenting or homogenising diverse views.<br><br>b) AI-driven micro-targeting fragments electoral constituencies into behavioural segments, undermining the collective character of representation.<br><br>c) AI advisory tools shaping legislative agenda-setting or speechwriting, introduce further delegation between representatives and their constituents. | 1.**AI tool adoption surveys**: proportion of legislators/political parties using AI for constituency analysis, drafting or communication (survey-based, e.g. 2024 study by the Inter-Parliamentary Union)<br>2. **Responsiveness gap metrics**: correlation between AI-segmented public opinion data and legislative voting records versus direct constituency surveys (sources: national election studies, public voting records)<br>3. **Representational equity**: whether AI tools systematically advantage certain demographic groups (e.g. digital access inequalities skewing AI-derived "public opinion") (sources: national or EU digital-inclusion data; e.g Eurostat or Ofcom)<br>4. **Disclosure requirements**: legislative or party-level rules mandating disclosure of AI use in representative functions; cross-country coding of rules (sources: OECD.AI data) |
| | *Delegating campaign operations to AI vendors/consultants by parties and candidates creates information asymmetries:*<br><br>a) Vendors possess proprietary knowledge of algorithmic tools that parties cannot audit.<br><br>b) AI lowers barriers for some new actors while raising them for others lacking access to data infrastructure, potentially reshaping competitive dynamics.<br><br>c) AI-driven disinformation operations and deep-fake deployment is used to destabilise competitors, raising the cost of honest competition. | 1. **Campaign AI expenditure**: party/campaign spending on AI tools and consultant services as a share of total campaign expenditure (sources: election finance disclosure data)<br>2. **In-house technical capacity**: survey measure of whether parties possess internal expertise to audit algorithmic tools, or are fully dependent on vendors (sources: surveys, requires primary research)<br>3. **Electoral volatility and fragmentation**: party system fragmentation indices (e.g. effective number of parties) over time, tested for correlation with AI adoption rates (sources: ParlGov database)<br>4. **Disinformation incidents**: documented AI-generated disinformation events per electoral cycle (e.g. EU Disinfo Lab, EUvsDisinfo (EEAS), Stanford Cyber Policy Center databases)<br>5. **Regulatory variation**: comparison of data protection and political advertising transparency rules across jurisdictions and their |

| Domain of AI Use | Delegation Mechanism | Empirical Measurement Indicators |
|---|---|---|
| | d) Differential access to AI capability may entrench incumbency advantages. | effect on competitive dynamics (e.g. CEPS's AI World databases) |
| **Public Administration** | *Delegating decisions to AI systems interrupts the traditional accountability chain from citizen to legislator to bureaucrat:*<br>a) The deploying agency may be unable to reconstruct the reasoning behind a specific output.<br>b) Corporate AI vendors introduce further opacity, with proprietary models shielded by trade secrecy.<br>c) Citizens and oversight bodies cannot link policy outcomes to identifiable human deciders. | 1. **Algorithmic opacity index**: results of Freedom of Information-type requests testing whether agencies can explain AI-reached decisions; refusal rates and quality of explanations (Sources: national public repositories)<br>2. **Explainability compliance**: proportion of public-sector AI deployments meeting explainability standards under applicable law (e.g. GDPR, EU AI Act) (sources: nazyional Algorithmic Transparency Records, such as ATRS in UK)<br>3. **Audit availability**: existence and independence of algorithmic audits; publication rates of audit results (sources: national audit offices, e.g. CNIL in France or NAO in UK)<br>4. **Vendor contract transparency**: public availability of procurement contracts for AI systems; extent of trade-secrecy carve-outs (Sources: public tender & procurement data, e.g. TED for EU tenders or BOSA for Belgium procurement data)<br>5. **Parliamentary scrutiny**: frequency and quality of legislative oversight hearings on AI in government; existence of dedicated AI oversight bodies (sources: national parliaments' debate/scrutiny records, e.g. OPECST in France) |
| | *AI intermediation distorts each stage of the responsiveness chain, undermining representatives' ability to accurately perceive, process, and act on citizen preferences:*<br>a) Preference-detection tools (sentiment analysis, social listening AI) may systematically misread public opinion by over-representing digitally active or data-rich demographics or misclassifying nuanced positions.<br>b) AI systems deployed for service delivery delegate discretionary judgement to algorithmic rules, potentially | 1. **Preference-detection accuracy**: validation studies comparing AI sentiment/opinion tools against survey benchmarks; demographic bias audits of AI-derived opinion data (Sources: national election studies, European Commission's JRC and Text Mining & Analysis Competence Centre studies)<br>2. **Policy congruence measures**: standard congruence methodology testing whether AI-mediated preference aggregation improves or degrades alignment between citizen preferences and policy outputs (Sources: Manifesto project + national election studies + sentiment analysis)<br>3. **Administrative discretion loss**: case-study and (fidelity of implementation) FOI-based assessment of proportion of administrative decisions where AI system output was determinative versus advisory (Sources:<br>4. **Procurement dependencies**: proportion of critical administrative functions delegated to |

| Domain of AI Use | Delegation Mechanism | Empirical Measurement Indicators |
|---|---|---|
| | reducing adaptive responsiveness to individual circumstances.<br><br>c) AI-enabled personalisation can segment the public into incommensurable preference-niches, making aggregate responsiveness difficult to define or measure.<br><br>d) The trustee/delegate relationship may be distorted as AI advisory systems nudge officials toward data-optimised rather than normatively deliberated positions. | external AI vendors versus publicly developed infrastructure (Sources: national FOI + ombudsman cases databases)<br>5. **Capacity asymmetries**: comparative measure of in-house government technical expertise versus sophistication of deployed AI systems; staff-to-AI-system ratios in agencies (Sources: EU Digital Economy and Society Index (DESI) data, national skills/workforce evidence, e.g. NAO in UK)<br>6. **Citizen satisfaction**: longitudinal surveys on perceived government responsiveness in AI-intensive versus non-AI-intensive service contexts (Sources: ESS, ESS, Eurobarometer, OECD Trust Survey, but AI-intensity of a public service is not yet an available variable in most cases) |

*Source: Authors' own elaboration.*

As the sources and empirical fieldwork examples listed in Table 1 show, most of the data that would be needed to assess in practice the dynamics of agency and control loss in delegation mechanisms when using AI in the selected main democratic domains is publicly available across EU countries and main advanced democracies. It can be easily collected by human-led, intensive desk research, or available upon request to the relevant public bodies to access existing databases. There are few important exceptions, of course, regarding the data that needs to be collected via dedicated surveys or that requires access to sensitive data or commercial information.

Overall, the indicators listed in Table 1 allows to measure the risks and potential failures identified in Figure 1. The integration of AI into democratic processes could three interlocking delegation failures: a monitoring failure, in which principals lose the capacity to observe how their AI agents actually behave; a failure of output contestability, in which the principals can no longer effectively challenge the decisions the agents produce; and goal misalignment, in which the agents' operative objectives diverge from the democratic values they were meant to serve. The significance of these failures lies more in their cumulative and mutually reinforcing character, since opacity prevents contestation and uncontested outputs in turn allow misalignment to consolidate, so that the accountability mechanisms on which representative democracy depends are progressively hollowed out even while formal institutions remain outwardly intact. Taken together, they imply that safeguarding democratic agency in an AI-mediated polity requires institutional designs that re-establish observability, contestability, and value alignment as enforceable conditions of delegation.

To explore further the risks entailed by these potential failures, in Table 2 we operationalise the NIST AI Risk Management Framework (NAIRMF) by translating its seven characteristics of trustworthy AI (valid and reliable, safe, secure and resilient, accountable and transparent, explainable and interpretable, privacy-enhanced, and fair with harmful bias managed) into domain-specific evaluation criteria for delegated AI systems in information ecosystems, elections, and public administration. Rather than treating trustworthiness as an abstract principle, the framework specifies observable indicators and governance requirements through which the performance of AI agents can be assessed when democratic principals delegate political, informational, or administrative functions to them. In doing so, it complements the principal–agent perspective by defining the substantive standards against which delegated AI systems should be evaluated in democratic contexts.

Integrating PAT with the NAIRMF is theoretically motivated, not ad hoc. Agency theory already treats risk and outcome uncertainty as internal to delegation rather than external to it (Eisenhardt 1989), so pairing it with a risk-management instrument simply extends one logic to where PAT stays silent, that is, the substance of what is delegated. PAT identifies where democratic delegation is vulnerable, e.g., information asymmetry, preference divergence, and weak monitoring and sanctioning (Eisenhardt 1989; Miller 2005), vulnerabilities that sharpen for highly capable agents through the competence–control dilemma (Abbott et al. 2020). The NIST trustworthiness characteristics then specify what principals actually need to assess once an agent acts on their behalf (NIST 2023).

Methodologically, risk-management frameworks are built to turn diffuse governance concerns into structured, auditable criteria (Black and Baldwin 2010), and the NAIRMF is exactly that kind of grid (NIST 2023). It converts each point of exposure into domain-specific criteria with measurable indicators, following NIST's map–measure–manage–govern logic.

Empirically, this makes the algorithmic "accountability gap" observable and comparable across domains. PAT yields indicators of delegation exposure—monitoring capacity, sanctioning mechanisms, information asymmetries—while the NAIRMF yields indicators of agent performance, operationalized through documented audit and assessment procedures (Raji et al. 2020). Together they enable comparison across democratic domains by asking not only whether AI systems affect participation, representation, or public oversight, but whether institutions can still monitor, contest, correct, and justify AI-mediated decisions. The pairing thus turns a normative concern into an empirical one: do democratic principals have the institutional capacity to evaluate, against standard trustworthiness criteria, the agents acting on their behalf?

**Table 2**: Trustworthy-AI evaluation criteria for delegated AI systems across democratic domains

| Domain of AI Use | Delegation Mechanism | NIST Trustworthiness Evaluation Criteria |
|---|---|---|
| **Information ecosystems** | *Citizens delegate information processing to AI-driven platforms (algorithmic curation, generative-AI summaries); platform operators and AI vendors act as agents.* | 1**. Valid and reliable:** factual accuracy of AI-generated summaries and recommended content; rate of hallucinated outputs (sources: EDMO, OECD AI Incidents Monitor)<br>2. **Safe:** prevalence of content harmful to public discourse (harassment, coordinated inauthentic behaviour) (sources: DSA Transparency Database, national public reporting portals for illegal/harmful online content, e.g. VIGINUM/PHAROS in France or Ofcom in UK, and platoforms' Adversarial Threat Reports)<br>3. **Secure and resilient:** resistance to data poisoning, prompt injection, and coordinated manipulation; service continuity under attack (sources: national cyber-security threat reporting, but most data are proprietary/non-public)<br>4. **Accountable and transparent:** disclosure of curation and ranking logic; quality of platform transparency reports; access for independent auditors (sources: DSA Transparency Database, EC transparency-report repository, DSA data-access (Art. 40) portal)<br>5. **Explainable and interpretable:** ability of users to understand why content was recommended or synthesised; availability of source provenance and attribution (sources: C2PA / Content Credentials adoption data, Ofcom research, targeted opinion surveys)<br>6**. Privacy-enhanced:** minimisation of behavioural data used for personalisation; controls over inferred profiles; lawful basis for processing (sources: GDPR Enforcement Tracker)<br>7. **Fair – with harmful bias managed:** demographic representativeness of sources; absence of systematic over- or under-exposure of viewpoints; routine bias audits (sources: national reports and casework on algorithmic discrimination, e.g. PEReN in France or Reuters Institute DNR studies) |
| **Elections** | *Parties, candidates, and elected representatives delegate voter analytics, targeting, and operational tasks to AI vendors and consultants (agents).* | 1. **Valid and reliable:** predictive accuracy of voter-behaviour and sentiment models against human data; documented validation procedures (sources: voter-model accuracy/validation is proprietary to campaigns and vendors; this is a known evidence gap)<br>2. **Safe:** protection against voter intimidation and synthetic-identity campaigns; incident-reporting procedures (sources:national election-period |

| Domain of AI Use | Delegation Mechanism | NIST Trustworthiness Evaluation Criteria |
|---|---|---|
| | | influence-operation detection and reports, OECD AIM / AIAAIC / AIID reports)<br>3. **Secure and resilient:** protection of voter and donor data; resilience of registration and electoral-management systems against adversarial attacks (sources: national electoral commissions/EMB data; IDEA international reports)<br>4. **Accountable and transparent:** disclosure of AI use in campaigns and political advertising; vendor-contract transparency (sources: Meta Ad Library + Google Ads Transparency outside EU, EU Political Advertising Regulation - TTPA repositories)<br>5. **Explainable and interpretable:** explainability of micro-targeting and ad-delivery decisions (sources: TTPA transparency notices, Meta Ad Library targeting data, volunteer browser data collection projects e.g. WhoTargetsMe)<br>6. **Privacy-enhanced:** lawful basis for voter-data processing (e.g., GDPR compliance); data minimisation in voter modelling; controls over data-broker inputs (sources: GDPR Enforcement Tracker, targeted studies)<br>7. **Fair – with harmful bias managed:** equitable access to AI tools across parties (especially smaller parties); absence of demographic bias in voter modelling; protections against vote-suppression effects (sources: research on fairness in digital campaigning such as Ada Lovelace Institute reports in UK) |
| **Public Administration** | *Civil servants and elected officials delegate decision-making, service delivery, and analytical tasks to AI systems and their corporate providers (agents).* | 1. **Valid and reliable:** accuracy of public-sector AI in target tasks (e.g., risk scoring, eligibility); routine validation (sources: national public algorithmic registers such as ATRS in UK, incident repositories (OECD AIM/AIAAIC/AIID))<br>2. **Safe:** assessment of harms to citizens from automated decisions in public sector (e.g., wrongful denials); availability of human-in-the-loop review; safe-deployment protocols (sources: OECD AIM / AIAAIC / AIID repositories).<br>3. **Secure and resilient:** cybersecurity standards for public-sector AI; resilience of critical service infrastructure (sources: national security standards and incident reports, e.g. ANSSI/CNIL repositories in France or ATRS in UK)<br>4. **Accountable and transparent:** documented decision rules; vendor-contract transparency; freedom-of-information compliance (sources: access-to-administrative-documents body/decisions repositories; e.g. CADA in France or Contracts Finder + ICO FOI decisions in UK)<br>5. **Explainable and interpretable:** capacity of agencies and affected citizens to obtain meaningful reasons for adverse decisions; availability of plain-language explanations (sources: legal/standards & |

| Domain of AI Use | Delegation Mechanism | NIST Trustworthiness Evaluation Criteria |
|---|---|---|
| | | complaint case-work repositories such as ATRS/ICO in UK, but lack of aggregate observational data on explanation quality)<br>6. **Privacy-enhanced:** lawful processing of citizen data; data minimisation; adoption of privacy-enhancing technologies (PETs) (sources: GDPR Enforcement Tracke)<br>**7. Fair – with harmful bias managed:** bias audits for protected groups; monitoring of differential error rates; disparate-impact assessment (sources: published impact audits and algorithmic-discrimination reports, e.g. DSIT bias research in UK, but they remain rare) |

*Source: Authors' own elaboration, based on NIST AI Risk Management Framework (NIST 2023).*

## 6. Conclusion

Our main contribution is a guide for identifying the factors researchers and policymakers should consider when assessing AI in political settings, especially where decisions depend on clear priorities and a defensible understanding of risks. Policies on the use or regulation of AI in politics, as well as preparations involving safeguards and countermeasures, often require impact assessments. Those assessments, however, need to be anchored in explicit metrics. Our proposed principal–agent perspective, grounded in established scholarship on democratic delegation, identifies where accountability is most exposed, while the NIST trustworthiness characteristics specify what principals should evaluate at those points, translating diffuse concerns about AI into standardized, auditable criteria. Neither element suffices alone: delegation analysis without substantive evaluation criteria allows for a diagnosis, but with limited measurement, and trustworthiness checklists without a theory of delegation allows for more fine-grained measurement, but without political meaning. Their integration converts democratic control over AI into an empirically assessable condition, namely institutional assessability, and thus offers a valid and empirically feasible entry point for structuring both evaluation research and policy practice.

Overall, it is crucial that future research address cascading principal-agent failures in AI governance, as the proposed analytical framework suggests. Traditional democratic accountability assumes traceable chains of delegation from citizens to representatives to bureaucrats. AI introduces cascading complexity: citizens cannot effectively monitor elected officials, who struggle to oversee agencies, which in turn cannot fully audit contractors, who themselves may be unable to explain algorithmic systems capable of exhibiting emergent behaviours beyond any human designer's intent. This is not only a cascade of delegation failures, but also a cascade of assessment failures: at each link, the capacity to evaluate the agent against trustworthiness criteria erodes further, so that the substantive standards exist on paper while the institutional capacity to apply them thins out along the chain. This creates principal-agent problems in every domain, intensifying opacity exponentially.

Future research should investigate whether democracy requires new institutional forms, such as “algorithmic ombudsmen” or mandatory interpretability intermediaries, that can bridge these accountability gaps. Additionally, scholars must examine whether AI’s capacity for autonomous adaptation fundamentally breaks the principal-agent model itself, as systems evolve beyond their initial mandates in ways that defy traditional oversight. Moreover, trustworthiness assessment, originally conceived as a one-off certification, fails for agents whose behaviour changes after evaluation, suggesting that continuous, policy cycle-based risk management may need to replace point-in-time auditing as the operational core of AI governance in democratic processes. Can democracies govern agents that learn, adapt, and potentially develop objectives misaligned with any human principal’s intent, and can they build institutions capable of continuously assessing whether such agents remain trustworthy at all?

## References

Abbott, K. W., Genschel, P., Snidal, D., & Zangl, B. (2020). Competence versus control: The governor's dilemma. *Regulation & Governance*, 14(4), 619–636. https://doi.org/10.1111/rego.12234

Achler, M., Krimmer, R., Kužel, R. O., Licht, N., & Rabitsch, A. (2022). Elections in digital times: A guide for electoral practitioners. UNESCO Publishing.

Altay, S., & Gilardi, F. (2024). People are skeptical of headlines labeled as AI-generated, even if true or human-made, because they assume full AI automation. *PNAS nexus*, *3*(10), pgae403.

Altman, D. (2026). The AI Democracy Dilemma. *Journal of Democracy*, *37*(1), 32-44.

Araujo, T. B., de Vreese, C. H., Helberger, N., Kruikemeier, S., van Weert, J. C. M., Bol, N., ... & Taylor, L. (2018). Automated decision-making fairness in an AI-driven world: Public perceptions, hopes and concerns. University of Amsterdam, Digital Communication Methods Lab.

Ash, E., Galletta, S., & Opocher, G. (2025). BallotBot: Can AI Strengthen Democracy?. Available at SSRN.

Borch, C. (2022). Machine learning, knowledge risk, and principal-agent problems in automated trading. *Technology in society*, *68*, 101852.

Budak, C., Nyhan, B., Rothschild, D. M., Thorson, E., & Watts, D. J. (2024). Misunderstanding the harms of online misinformation. *Nature*, *630*(8015), 45-53.

Busuioc, M. (2021). Accountable artificial intelligence: Holding algorithms to account. Public administration review, 81(5), 825-836.

Chapekis, A., & Lieb, A. (2025). Google users are less likely to click on links when an AI summary appears in the results. https://www.pewresearch.org/short-reads/2025/07/22/google-users-are-less-likely-to-click-on-links-when-an-ai-summary-appears-in-the-results/?gad_source=1&gad_campaignid=22378837192&gbraid=0AAAAA-ddO9HTjFbn9b487nH8fDBLQxI84&gclid=Cj0KCQiA2bTNBhDjARIsAK89wlF

9EYvr2YldImtfJr9UjPyo6S0Lp87BSgwYDUWrrGi2lybhdkYp6koaAuT_EALw_wcB

Coeckelbergh, M. (2024). Why AI undermines democracy and what to do about it. John Wiley & Sons.

Cofone, Ignacio and Strandburg, Katherine J., Algorithmic Opacity as a Principal-Agent Problem (February 16, 2026). 15 NYU Journal of Intellectual Property and Entertainment Law 1 (2026), Available at SSRN: https://ssrn.com/abstract=6249838 or http://dx.doi.org/10.2139/ssrn.6249838

Cools, H., & Diakopoulos, N. (2026). Uses of generative AI in the newsroom: Mapping journalists' perceptions of perils and possibilities. Journalism Practice, 20(3), 878-896.

Coppedge, M., Gerring, J., Altman, D., Bernhard, M., Fish, S., Hicken, A., ... & Teorell, J. (2011). Conceptualizing and measuring democracy: A new approach. Perspectives on politics, 9(2), 247-267.

Dahl, R. A. (1971). Polyarchy: Participation and opposition. Yale University Press.

Diamond, L., & Morlino, L. (2004). The quality of democracy: An overview. Journal of democracy, 15(4), 20-31.

Eady, G., Paskhalis, T., Zilinsky, J., Bonneau, R., Nagler, J., & Tucker, J. A. (2023). Exposure to the Russian Internet Research Agency foreign influence campaign on Twitter in the 2016 US election and its relationship to attitudes and voting behavior. *Nature communications*, *14*(1), 62.

Eisenhardt, K. M. (1989). Agency Theory: An Assessment and Review. *Academy of Management Review*, 14(1), 57–74. https://doi.org/10.5465/amr.1989.4279003

Estevez Almenzar, M., Fernandez Llorca, D., Gomez, E., & Martinez Plumed, F. (2022). Glossary of human-centric artificial intelligence (No. JRC129614). Joint Research Centre. https://publications.jrc.ec.europa.eu/repository/bitstream/JRC129614/JRC129614_01.pdf

European External Action Service. (2025, March). 3rd EEAS report on foreign information manipulation and interference threats: Exposing the architecture of FIMI operations. European Union. EEAS-3nd-ThreatReport-March-2025-05-Digital-HD.pdf

Farooq, A., & de Vreese, C. (2025). Deciphering authenticity in the age of AI: how AI-generated disinformation images and AI detection tools influence judgements of authenticity. AI & SOCIETY, 1-12.

Farooq, A., & de Vreese, C. (2026). Disinformation and its Sociopolitical Context. In Disinformation: A Multi-Disciplinary Analysis (pp. 3-22). Cham: Springer Nature Switzerland.

Fink-Hafner, D. (2025). Alternatives to liberal democracy and the role of AI: The case of elections. Politics and Governance, 13.

Fischhoff, B., Slovic, P., Lichtenstein, S., Read, S., & Combs, B. (1978). How safe is safe enough? A psychometric study of attitudes towards technological risks and benefits. *Policy sciences*, *9*(2), 127-152.

Floridi, L., & Cowls, J. (2022). A unified framework of five principles for AI in society. Machine learning and the city: Applications in architecture and urban design, S. Carta (Ed.), pp. 535-545, https://doi.org/10.1002/9781119815075.ch45 .

Foos, F. (2024). The use of AI by election campaigns. LSE Public Policy Review, 3(3).

Gillespie, T. (2024).Generative AI and the politics of visibility. *Big Data & Society,* 11, 1–14.

Grossman, S. J., & Hart, O. D. (1983). An Analysis of the Principal-Agent Problem, Econometrica, vol. 51, no. 1, January 1983, pp. 7-45.

Hackenburg, K., & Margetts, H. (2024). Evaluating the persuasive influence of political microtargeting with large language models. Proceedings of the National Academy of Sciences, 121(24), e2403116121.

Hackenburg, K., Tappin, B. M., Hewitt, L., Saunders, E., Black, S., Lin, H., ... & Summerfield, C. (2025). The levers of political persuasion with conversational artificial intelligence. Science, 390(6777), eaea3884.

Hansson, S. O. (2010). Risk: objective or subjective, facts or values. *Journal of Risk Research*, *13*(2), 231–238. https://doi.org/10.1080/13669870903126226

Heider, M., Stegherr, H., Nordsieck, R., & Hähner, J. (2023). Assessing model requirements for explainable AI: A template and exemplary case study. Artificial life, 29(4), 468-486.

Helberger, N., Pierson, J., & Poell, T. (2018). Governing online platforms: From contested to cooperative responsibility. The information society, 34(1), 1-14.

Hradec, J., Ostlaender, N., & Bernini, A. (2023). FABLES: Framework for Autonomous Behaviour-rich Language-driven Emotion-enabled Synthetic populations,https://publications.jrc.ec.europa.eu/repository/bitstream/JRC135070/JRC135070_01.pdf

Innerarity, D. (2024). The epistemic impossibility of an artificial intelligence take-over of democracy. AI & society, 39(4), 1667-1671.

Jobin, A., Ienca, M., & Vayena, E. (2019). The global landscape of AI ethics guidelines. Nature Machine Intelligence, 1(9), 389–399.

Almeida, M., Jungherr A., Sandri, G., Makhortykh, M., Galdon Clavell, G. & Rabitsch , A. (2026). EU Joint Research Centre (JRC), Science for policy report on: 'Impact of Artificial Intelligence (AI) on Democracy', forthcoming 2026.

Juneja, P. & International Institute for Democracy and Electoral Assistance. (2024). Artificial intelligence for electoral management. International IDEA. Artificial Intelligence for Electoral Management | International IDEA

Jungherr, A. (2023). Artificial intelligence and democracy: A conceptual framework. Social media+ society, 9(3), 20563051231186353.

Jungherr, A., & Rauchfleisch, A. (2025). Artificial Intelligence in deliberation: The AI penalty and the emergence of a new deliberative divide. Government Information Quarterly, 42(4), 102079.

Jungherr, A., & Schroeder, R. (2023). Artificial intelligence and the public arena. Communication Theory, 33(2-3), 164-173.

Jungherr, A., Rauchfleisch, A., & Wuttke, A. (2026). Artificial intelligence in election campaigns: Perceptions, penalties, and implications. *Political Communication*, 1-22, https://doi.org/10.1080/10584609.2025.2611913

Jungherr, A., Schroeder, R., & Stier, S. (2019). Digital media and the surge of political outsiders: Explaining the success of political challengers in the United States, Germany, and China. Social Media+ Society, 5(3), 2056305119875439.

Karppinen, K., Moe, H., & Svensson, J. (2025). What Democracy Are We Talking About? Reviewing the Debate on the Implications of Generative AI for Democracy. Journal of Information Policy, 15.

Kersting, N. (2012). The future of electronic democracy. In N. Kersting (Ed.), Electronic democracy (pp. 11–54). Verlag Barbara Budrich.

Kersting, N., & Baldersheim, H. (Eds.). (2004). Electronic voting and democracy: a comparative analysis. Springer.

König, P. D. (2025). Understanding the politics of artificial intelligence. Edward Elgar Publishing.

Koops, B. J., Hildebrandt, M., & Jaquet-Chiffelle, D. O. (2010). Bridging the accountability gap: Rights for new entities in the information society?. Minn. JL Sci. & Tech., 11, 497.

Kreps, S., & Kriner, D. (2023). How AI threatens democracy. Journal of Democracy, 34(4), 122-131.

Landemore, H. (2024). Can artificial intelligence bring deliberation to the masses? In R. Chang & A. Srinivasan (Eds.), Conversations in philosophy, law, and politics. Oxford University Press.

Lijphart 2012[1999]. Patterns of democracy: Government forms and performance in thirty-six countries (2nd ed.). Yale University Press.

Magin, M., Geiß, S., Stark, B., & Jürgens, P. (2021). Common core in danger? Personalized information and the fragmentation of the public agenda. *The International Journal of Press/Politics*, *27*(4), 887–909.

Makhortykh, M., De León, E., Urman, A., & Gil-López, T. (2025). Where did you come from, where did you go? News trajectories in Germany and Switzerland. *Journal of Computational Social Science*, *8*, 1–21

Makhortykh, M., Sydorova, M., Baghumyan, A., Vziatysheva, V., & Kuznetsova, E. (2024). Stochastic lies: How LLM-powered chatbots deal with Russian disinformation about the war in Ukraine. Harvard Kennedy School Misinformation Review.

Martin, Z., Jackson, D., Trauthig, I. K., & Woolley, S. C. (2024). *Political machines: Understanding the role of generative AI in the U.S. 2024 elections and beyond*. Center for Media Engagement, The University of Texas at Austin. Political Machines: Understanding the Role of AI in the U.S. 2024 Elections and Beyond - Center for Media Engagement - Center for Media Engagement

McDaniel, J. L., & Pease, K. (Eds.). (2021). Predictive policing and artificial intelligence. Routledge, Taylor & Francis Group.

Miller, G. J. (2005). The political evolution of principal-agent models. Annual Review of Political Science, 8, 203–225. https://doi.org/10.1146/annurev.polisci.8.082103.104840

Mittelstadt, B. D., Allo, P., Taddeo, M., Wachter, S., & Floridi, L. (2016). The ethics of algorithms: Mapping the debate. Big Data & Society, 3(2), 1–21.

Mollick, E. (2024). Co-intelligence: Living and working with AI. Portfolio/Penguin.

Moon, A., & Boudreaux, B. (2026). A formal model of how artificial intelligence erodes human agency (RR-A4817-1). RAND Corporation. https://doi.org/10.7249/RRA4817-1

Narayanan, A., & Kapoor, S. (2024). AI snake oil: What artificial intelligence can do, what it can't, and how to tell the difference. Princeton University Press.

National Institute of Standards and Technology (NIST). (2023). Artificial Intelligence Risk Management Framework (AI RMF 1.0). NIST AI 100-1. U.S. Department of Commerce. https://doi.org/10.6028/NIST.AI.100-1

Newman, J., Mintrom, M., & O'Neill, D. (2022). Digital technologies, artificial intelligence, and bureaucratic transformation. Futures, 136, 102886.

Nord, M., Altman, D., Angiolillo, F., Fernandes, T., Good God, A., & Lindberg, S. I. (2025). *Democracy report 2025: 25 years of autocratization – Democracy trumped?* V-Dem Institute, University of Gothenburg. https://www.v-dem.net/documents/60/V-dem-dr__2025_lowres.pdf

Norris, P. (2012). Making democratic governance work: How regimes shape prosperity, welfare, and peace. Cambridge University Press.

Novelli, C., & Sandri, G. (2025). Digital democracy. A Companion to Digital Ethics, 241-256.

Novelli, C., Casolari, F., Rotolo, A., Taddeo, M., & Floridi, L. (2024). AI risk assessment: A scenario-based, proportional methodology for the AI Act. Digital Society, 3(1), Article 13. https://doi.org/10.1007/s44206-024-00095-1

Novelli, C., Taddeo, M., & Floridi, L. (2024). Accountability in artificial intelligence: what it is and how it works. Ai & Society, 39(4), 1871-1882.

Novelli, C., Formisano, G., Juneja, P., Sandri, G., & Floridi, L. (2024). Artificial Intelligence for the Internal Democracy of Political Parties. Minds and Machines, 34(4), 36.

OECD (2025), *Governing with Artificial Intelligence: The State of Play and Way Forward in Core Government Functions*, OECD Publishing, Paris, https://doi.org/10.1787/795de142-en.

Ojanen, A., Björk, A., Singh, S., & Djakonoff, V. (2026). *A framework for democratic AI governance* (KT4D Deliverable D5.1). Demos Helsinki. https://demoshelsinki.fi/publication/a-framework-for-democratic-ai-governance/

Olejnik, L. (2025). AI Propaganda factories with language models. arXiv preprint arXiv:2508.20186.

Pahlka, J. (2023). Recoding America: why government is failing in the digital age and how we can do better. Metropolitan Books.

Persily, N., & Tucker, J. A. (Eds.). (2026). Artificial intelligence, politics, and political science. Cambridge University Press, https://apsanet.org/wp-content/uploads/2026/05/Lee-TF-APSA-AI-Report-2026-Tucker-Persily.pdf

Petrone, J. (2025). AI Leap 2025: Estonia sets the global standard for AI in education. e-Estonia. https://e-estonia.com/ai-leap-2025-estonia-sets-ai-standard-in-education/

Phelps, S., & Ranson, R. (2023). Of models and tin men: a behavioural economics study of principal-agent problems in AI alignment using large-language models. *arXiv preprint arXiv:2307.11137*.

Pi, Y., & Proctor, M. (2025, January). Toward empowering AI governance with redress mechanisms. In Cambridge Forum on AI: Law and Governance (Vol. 1, p. e24). Cambridge University Press.

Posnett K. (2025). The new markets for AI data. *Financial Times*. The new markets for AI data | Goldman Sachs

Power, S., & Jeong, G. (2026). Harnessing artificial intelligence to enhance political finance oversight. International IDEA. https://doi.org/10.31752/59539

Raji, I. D., Smart, A., White, R. N., Mitchell, M., Gebru, T., Hutchinson, B., Smith-Loud, J., Theron, D., & Barnes, P. (2020). Closing the AI accountability gap: Defining an end-to-end framework for internal algorithmic auditing. In Proceedings of the 2020 Conference on Fairness, Accountability, and Transparency (FAT* '20) (pp. 33–44). ACM.

Rizk, A., & Lindgren, I. (2025). Automated decision-making in public administration: Changing the decision space between public officials and citizens. Government Information Quarterly, 42(3), Article 102061

Saeri, A. K., Graham, J., Noetel, M., Slattery, P., Ah-king, D., Aittokallio, E., ... & Shen, T. (2026). Prioritization of Risks from Artificial Intelligence: A Delphi Study of 272 International Experts. arXiv preprint arXiv:2606.04490.

Salvi, F., Horta Ribeiro, M., Gallotti, R., & West, R. (2025). On the conversational persuasiveness of GPT-4. Nature Human Behaviour, 9, 1645–1653.

Sass, R., Novelli, C., & Zio, E. (2026). Quantifying values: The problem of AI risk. SSRN. https://doi.org/10.2139/ssrn.6140546

Scharre, P. (2018). Army of none: Autonomous weapons and the future of war. W. W. Norton & Company.

Scherer, T., Fecher, H., Jost, P., & Kruschinski, S. (2026). Balancing Innovation and Control: Public Views on Risks and Governance of AI in Political Communication. *Policy & Internet*, *18*(1), e70027.

Shapira, A., & Rahat, G. (2017). The Intra-Party Democracy Index: Development, Application, and Results 1. In The Elections in Israel 2013 (pp. 107-130). Routledge.

Simon, F. M. (2025). Rationalisation of the news: How AI reshapes and retools the gatekeeping processes of news organisations in the United Kingdom, United States and Germany. *New Media & Society*, 14614448251336423.

Simon, F. M., & Altay, S. (2025). Don't panic (yet): Assessing the evidence and discourse around generative AI and elections. szw12z757x

Skotnicka-Zasadzień, B., & Wolniak, R. (2025). *Artificial Intelligence in Public Administration*. Taylor & Francis.

Slattery, P., Saeri, A. K., Grundy, E. A. C., Graham, J., Noetel, M., Uuk, R., Dao, J., Pour, S., Casper, S., & Thompson, N. (2026). The AI risk repository: A meta-review, database, and taxonomy of risks from artificial intelligence. Patterns, 7(5), Article 101517. https://doi.org/10.1016/j.patter.2026.101517

Slovic, P. (1987). Perception of risk. *science*, *236*(4799), 280-285.

Sousa, I. A. D. N., Machado, J., & Vaz, J. C. (2025). Generative AI as a catalyst for democratic Innovation: Enhancing citizen engagement in participatory budgeting. *arXiv preprint arXiv:2509.19497*.

Strøm, K. (2000). Delegation and accountability in parliamentary democracies. *European Journal of Political Research*, 37(3), 261–290. https://doi.org/10.1111/1475-6765.00513

Suleyman M. & Bhaskar M. (2025).The coming wave: technology, power, and the twenty-first century's greatest dilemma. Crown.

Tangi, L., Müller, A. P. R., & Janssen, M. (2025). AI-augmented government transformation: Organisational transformation and the sociotechnical implications of artificial intelligence in public administrations. Government Information Quarterly, 42(3), 102055.

Toff, B., & Simon, F. M. (2025). "Or they could just not use it?": The dilemma of AI disclosure for audience trust in news. The International Journal of Press/Politics, 30(4), 881-903.

Tomini, L., & Sandri, G. (2018). The quality of democracy: Towards a new research agenda. In Challenges of Democracy in the 21st Century (pp. 1-14). Routledge.

Vaccari, C., & Chadwick, A. (2020). Deepfakes and disinformation: Exploring the impact of synthetic political video on deception, uncertainty, and trust in news. Social media+ society, 6(1), 2056305120903408.

Valentim, V., & Dinas, E. (2024). Does party-system fragmentation affect the quality of democracy?. British Journal of Political Science, 54(1), 152-178.

Van Noordt, C., & Misuraca, G. (2022). Artificial intelligence for the public sector: results of landscaping the use of AI in government across the European Union. *Government information quarterly*, *39*(3), 101714.

Votta, F., Dobber, T., Guinaudeau, B., Helberger, N., & de Vreese, C. (2025). The cost of reach: Testing the role of ad delivery algorithms in online political campaigns. *Political Communication*, *42*(3), 476-508.

Vries, C. E. D., & Hobolt, S. (2020). Political entrepreneurs: The rise of challenger parties in Europe. Princeton University Press.

Wolfs, W. (2025). *Political finance in the digital age: Towards evidence-based reforms*. International Institute for Democracy and Electoral Assistance (International IDEA). Political Finance in the Digital Age: Towards Evidence-Based Reforms | International IDEA

Wuttke, A., Rauchfleisch, A., & Jungherr, A. (2025). Artificial intelligence in government: Why people feel they lose control. *arXiv preprint arXiv:2505.01085*.